\PassOptionsToPackage{table}{xcolor}

\documentclass[acmtog]{acmart} 

\acmSubmissionID{2717} 
\AtBeginDocument{%
  }

\copyrightyear{2026}
\acmYear{2026}
\setcopyright{cc}
\setcctype{by}
\acmConference[SA Conference Papers '26]{SIGGRAPH Asia 2026 Conference Papers}{December 01--04, 2026}{Kuala Lumpur, Malaysia}
\acmBooktitle{SIGGRAPH Asia 2026 Conference Papers (SA Conference Papers '26), December 01--04, 2026, Kuala Lumpur, Malaysia}
\acmDOI{10.1145/3829340.3842373}
\acmISBN{979-8-4007-2842-6/2026/12}

\author{Patrick Attimont}
\email{patrick.attimont@inria.fr}
\affiliation{%
  \institution{INRIA - Grenoble University}
  \city{Grenoble}
  \country{France}}
\orcid{0009-0008-0450-6523}

\author{Kartic Subr}
\email{K.Subr@ed.ac.uk}
\affiliation{%
  \institution{University of Edinburgh}
  \city{Edinburgh}
  \country{United Kingdom}
}
\orcid{0000-0002-7302-4383}

\author{Cyril Soler}
\affiliation{%
  \institution{INRIA - Grenoble University}
  \city{Grenoble}
  \country{France}}
\email{cyril.soler@inria.fr}
\orcid{0000-0003-3214-4183}

\renewcommand{\shortauthors}{Attimont et al.}

\begin{abstract}
We present a novel method for computing global illumination by expressing the solution to the light transport equation as a 13D Gaussian mixture model
over positions, directions, surface normals, and material properties. We show that including scene properties in the Gaussian representation drastically reduces
the number of functions and speeds up evaluation. As opposed to traditional light transport methods based on Neumann series, the parameters of
our model are directly estimated by minimizing the residual of the rendering equation. 
While both optimization and rendering require repeated evaluations of a linear combination of high-dimensional Gaussian functions, 
we introduce an efficient culling strategy to keep the optimization tractable and produce renderings in real time. 
Our representation enables to render fast, view-independent solutions to the light transport equation, achieving rendering times on the 
order of milliseconds, with a fraction of the memory requirements of conventional neural rendering approaches.
\end{abstract}


  
\usepackage{graphicx} 
\usepackage{tikz}

\usepackage{enumitem}
\usepackage{color}
\usepackage[table]{xcolor} 
\usepackage{booktabs}      
\usepackage{siunitx}       
\usepackage{multirow}
\usepackage{calc}
\usepackage{colortbl}

\definecolor{myBlue}  {rgb}{0.302,0.345,0.631}
\definecolor{myGreen} {rgb}{0.125,0.651,0.329}
\definecolor{myRed}   {rgb}{0.651,0.125,0.329}
\definecolor{myGrey}  {rgb}{0.651,0.651,0.651}
\definecolor{myOrange}{rgb}{0.651,0.451,0.451}

\DeclareMathSymbol{\intercal}{\mathbin}{AMSa}{"7C}                                                                               

\newcommand{\myparagraph}[1]{\smallskip{\noindent\bf #1.}}

\definecolor{best}{rgb}{0.60,0.85,0.60}
\definecolor{second}{rgb}{0.82,0.94,0.82}
\definecolor{third}{rgb}{1.00,0.97,0.70}

\begin{teaserfigure}
  \centering
\includegraphics{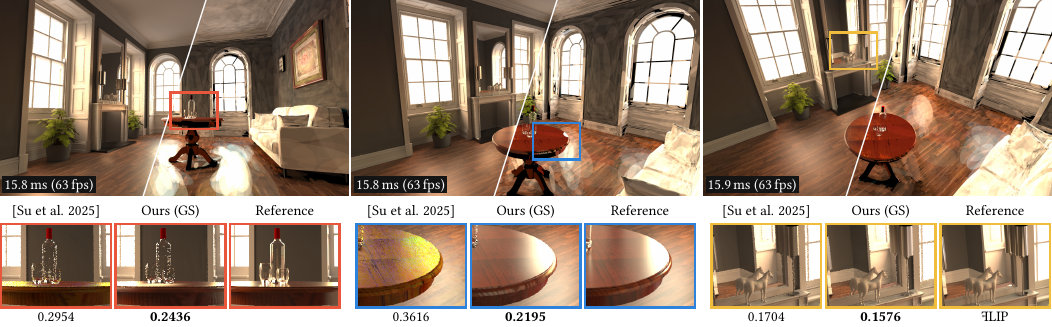}
  \caption{
        Our method represents global illumination as an explicit mixture of 13D Gaussians optimized directly against the residual of the rendering equation.
        On \textsc{Living Room}, we train for 11m30s and render from arbitrary viewpoints at 1920$\times$1080, 1~spp, 63 fps.
        We achieve lower error and render 3.7$\times$ faster than Vertex Features \cite{su_vertex_2025} which we train for 4h29m.
        Each image above is split between the rendered Gaussians and their flattened visualization along the scene geometry.
    }
  \Description{~}
  \label{fig:teaser}
\end{teaserfigure}

\usepackage{subcaption}
\usepackage{rotating}
\usepackage{caption}
\usepackage{array}

\title{Gaussian Light Transport}

\begin{document}
\maketitle

\section{Introduction}
Efficient computation of global illumination in 3D environments is a long-standing problem in computer graphics. Solving the integral equation~\cite{kajiya_rendering_1986}---which models the equilibrium of directional radiant light energy---would yield a radiance field which can be used to render photo-realistic images from any viewpoint. This remains an elusive goal, due to the recursive nature of the equation which unravels as infinite-dimensional integrals. In this paper, we represent the solution to the equation as a sum of Gaussians and optimize the parameters of the Gaussians to approximately satisfy the equation.  

Previous approaches have uncovered trade-offs between \emph{View-Independent Rendering} by solving the global radiance field vis-a-vis view-dependent solutions. Galerkin methods (such as radiosity~\cite{goral_modeling_1984}) yield global solutions but are impractical for representing high-frequency effects due to reflections, caustics, shadows, etc. On the other hand, naive path tracing~\cite{kajiya_rendering_1986} tends to require heavy computation for smooth effects due to global illumination. The \emph{de facto} approach for contemporary light transport solutions is Monte Carlo integration of the Neumann series expansion of the rendering equation, yielding a general and effective solution. View-dependent methods require recomputation for novel views within a scene, but are usually justified when there are dynamic elements. We propose a method for view-independent rendering that is effective for multi-view rendering (e.g. fly-throughs) of complex, static scenes. 
In other words, ``view-independent'' means that the calculation is not tied to the parameters of a camera (as path tracing would),
and computes the entire light field that solves the transport equation. View-dependent images are thereafter produced by 
slicing this light field for a given camera.

Several methods from machine learning have been adapted for rendering~\cite{MLRendercourse2018}. Image-based rendering approaches such as Neural Radiance Fields (NeRFs)~\cite{mildenhall_neural_2020} and Gaussian Splatting~\cite{kerbl_gaussian_2023} aim to hypothesize arbitrary views given multi-view images as input. Neural radiosity optimizes a neural field, defined on a discrete grid, to minimize the residual of the rendering equation~\cite{hadadan_neural_2021,su_dynamic_2024}.

\begin{figure*}[htbp]
	\centering
	\includegraphics[width=1.0\textwidth]{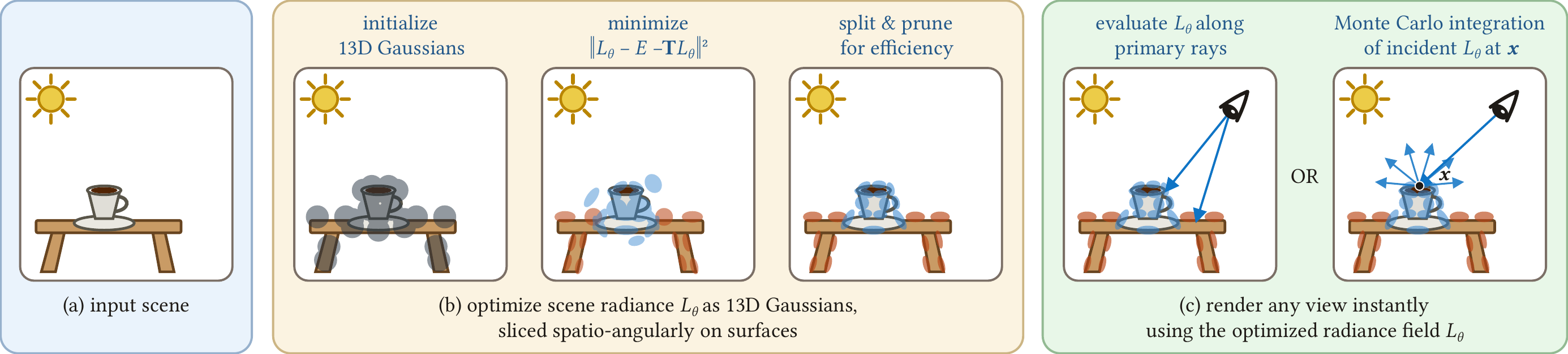}
	\caption{
	Flatland illustration of our method.
	We represent the solution to the light transport equation as a sum of $13$D Gaussians over positions, directions, normals, and material properties---only the spatial slice is shown.
	Given an input scene (a), we (b) optimize the global scene radiance and (c) render multi-view images from it.
	Gaussians are initialized on scene surfaces and fitted to minimize the residual of the rendering equation.
	We render novel views either by evaluating the optimized radiance at the primary hit (c. left)
 	or by tracing secondary rays and using the model as the incident radiance at $\mathbf {x}$ (c. right).
	Whether fitting or rendering, the model is only ever queried at ray-traced points, so only the first visible surface contributes and visibility needs no explicit treatment.}
	\label{fig:overview}
\end{figure*}

We propose a new view-independent method for solving the light transport
equation by leveraging the versatile and adaptive nature of Gaussian kernels.
We represent the solution to the light transport equation as the sum of 13D
Gaussians over the full spaces of positions, directions, normals, and material properties,
and optimize these parameters to minimize the residual of the light transport
equation. 
Our Gaussian kernels are therefore not restricted to remain on surfaces and inherently
handle local variations in normals and materials. This allows us to compute a
view-independent approximation to the solution to the rendering equation at a
fraction of the memory cost of neural methods. We exploit the local support of
Gaussians to design a culling strategy for a fast evaluation of the model,
enabling efficient optimization (in a matter of minutes) and real-time rendering (see Figure~\ref{fig:overview}).  In summary,
\begin{itemize}
	\item we propose a novel approach to solving the light transport equation by directly optimizing Gaussians, to obtain a view-independent solution;
	\item the complexity of our model adapts to the solution and is not constrained by the geometry of the scene;
	\item we show that this approach matches or improves on existing neural methods in quality, while training \num{10}--\num{23}$\times$ faster and rendering \num{2}--\num{4}$\times$ faster, with a memory footprint of just a few megabytes.
\end{itemize}

\section{Related Work}


\subsection{Global illumination methods}

\myparagraph{Galerkin and Collocation Methods}
Galerkin radiosity~\cite{goral_modeling_1984,zatz_galerkin_1993} and directional finite element 
methods~\cite{heckbert_finite_1991} project solutions onto structured basis
functions defined over the scene geometry. They involve solving a linear
system and provide view-independent solutions with theoretical guarantees.
However, the need to parameterize geometry and directions makes them poorly suited to complex scenes with intricate geometry or
highly glossy materials. Representing radiance as a 5D function leads to a prohibitive memory cost for fine basis functions that are required to capture phenomena such as caustics and glossy reflections. Collocation methods, including meshless formulations~\cite{lehtinen_meshless_2008,christensen_point-based_2008,wang_wavelet_2015}, alleviate some discretization constraints but do not fundamentally resolve these scalability issues.

\myparagraph{Monte Carlo and Path Tracing Methods}
Monte Carlo methods avoid explicit function representations by estimating the
solution through stochastic sampling, typically via 
a Neumann series expansion of the rendering equation. Path tracing and its
extensions---including bidirectional path tracing~\cite{veach_robust_1997},
photon mapping~\cite{jensen_global_1996,hachisuka_progressive_2008}, and
Metropolis Light Transport~\cite{veach_metropolis_1997}---are highly effective at
handling complex light transport phenomena. They are flexible, but can exhibit noise from high variance when long light paths are involved,
making them computationally expensive and poorly suited to view-independent global illumination.
Path re-using methods~\cite{ouyang_restir_2021}, temporal reprojection~\cite{schied_spatiotemporal_2017} and gradient-domain path tracing~\cite{kettunen_gradient_2015} 
mitigate this problem by making the best of already computed sub-parts of the light transport for new views in their own specific ways.

One other limitation of path tracing methods is their inherent ignorance of the spatial and angular bandwidth of the reconstructed signal,
causing high variance in low frequency parts of the image.  Frequency-aware splatting~\cite{belcour_covariance_2013} and adaptive low-frequency 
methods~\cite{jarosh_progressive_2011} partially address these issues. 

\subsection{View-independent precomputed radiance}

\myparagraph{PRT and Radiance caching}
Precomputed Radiance Transfer (PRT) methods compute light transport matrices that allow 
scene elements such as view points and
light~\cite{sloan_precomputed_2002,ng_triple_2004}, scene
geometry~\cite{pan_precomputed_2007} and materials~\cite{sun_interactive_2007} 
to be changed interactively when 
rendering novel views.  More recently, \citet{guo_precomputed_2024} combine precomputed radiance transport
with 3D Gaussian splatting to interactively generate images with indirect
illumination---as opposed to global illumination.  PRT methods struggle to
locally adapt to the radiance field, leading to heavy memory requirements. This is further exacerbated by the basis functions (wavelets, spherical harmonics, etc.) being tied to the
geometry.

Radiance caching (and irradiance caching)~\cite{krivanek_radiance_2008} accelerates global illumination for non-Lambertian scenes by precomputed storage of radiance in a world-space
grid. This can be stored on surfaces~\cite{tatzgern_radiance_2024} or in a volumetric grid~\cite{xing_realtime_2024}. 
Novel views are rendered by simply convolving the BRDFs with incoming radiance from the cache. Storage requirements of these methods are high (gigabytes) but rendering can be an order of magnitude faster (10 ms) than PRT. 
These methods make use of their precomputed cache to retrieve indirect illumination only,
while our evaluated radiance fields include both direct and indirect illumination as a proxy.

\myparagraph{Alternative representations}
Several approaches trade physical rigor for efficiency and compactness. Data-driven
bases~\cite{belcour_data_2022} model the space of solutions using learned
low-dimensional representations, enabling efficient relighting. Point-based and volumetric approaches such as point-based radiance fields~\cite{zhang_differentiable_2023}, Plenoxels~\cite{yu_plenoxels_2021},
and Gaussian splatting~\cite{kerbl_gaussian_2023} optimize for visual metrics. Neural radiance fields
(NeRF)~\cite{mildenhall_neural_2020} learn continuous volumetric representations of radiance from images and render novel views via ray integration. 
Recent methods use 3D Gaussian Splatting for radiance caching~\cite{bauer_gscache_2026}, and Gaussian mixtures for path guiding~\cite{vorba_path_2019}.
\citet{diolatzis_n_dimensional_2024} fit rendered light fields with Gaussians, conditioned on G-Buffers, using geometry, material, and normalized scene-variable information to efficiently model high-quality, view- and scene-dependent appearance. 
These approaches have triggered numerous extensions including for dynamic
scenes~\cite{coomans_realtime_2024}. While highly effective for view synthesis, these methods typically approximate observed radiance distributions (i.e. pre-acquired or pre-rendered data) and are not physically based nor solve the rendering equation.

\myparagraph{Neural Approaches}
Neural radiosity~\cite{hadadan_neural_2021} represents radiance using
a neural network trained to minimize the residual of the rendering equation over a 
dense grid of points that is unaware of the frequency content of the represented light field.
Its extension to dynamic scenes has also been proposed~\cite{su_dynamic_2024}.
A key limitation of these approaches is their lack of adaptivity to scene
content, resulting in a large memory footprint due to uniform
discretizations (e.g., 5D grids). Recent work mitigates this issue by attaching
neural representations to the scene geometry~\cite{su_vertex_2025}, improving
efficiency but still tying the representation to the underlying mesh and material properties,
or by using multiresolution hash encoding~\cite{muller_instant_2022}.

\section{Technical Description}

Our goal is to optimize a radiance field to solve the rendering equation in a synthetic scene.
The equilibrium of light transport in a scene is governed by the rendering equation~\cite{kajiya_rendering_1986}
		\begin{align} L(\mathbf x,\omega) = E(\mathbf x,\omega) + \int_\Omega  L_i (\mathbf x,\omega')f(\mathbf x,\omega,\omega') \mathbf n \cdot \omega' d\omega',\label{eqn:kajiya}\end{align}
where $\Omega$ is the upper hemisphere of directions, $L_i(\mathbf x,\omega')$ is the incident radiance at  $\mathbf x$ along $\omega'$, $f$ is the reflectance function, $\mathbf n$ is the normal at $\mathbf x$, and $E$ is emitted radiance. In operator form, with the linear light transport operator $\mathbf T$, this equation is~\cite{arvo_role_1995}:
			\begin{align} L = E + \mathbf TL.\label{eqn:rendering}\end{align}

We seek to solve Equation~\ref{eqn:rendering} with a  
parameterized radiance function $\theta\mapsto L_\theta$ with $\theta\in\mathbb R^n$, by minimizing its residual:
			$$\theta^* = \mbox{argmin}_\theta \| r_\theta \|^2 \mbox{~~~~~~~\;\;with\;\;~~~~~~~} r_\theta=L_\theta - E - \mathbf T L_\theta. $$
In contrast to the Neural Radiosity framework that represents $L$ using a neural network~\cite{hadadan_neural_2021}, we use an explicit sum of localized Gaussian kernels, which allows adapting the location and number of functions to the spatial and angular frequency content of the solution $L_{\theta^*}$ leading to lower memory requirements.

Our training objective is the squared $L_2$ norm of the residual of the rendering equation over the 
surface area and angular domains:
\begin{align}
	\mathcal{L}(\theta)&=\|r_\theta\|^2 = \int_\mathcal{S}\int_\Omega r_\theta(\mathbf x, \omega)^2\; d\omega d\mathbf x.\label{eqn:loss}
\end{align}
We compute $L_\theta$ by combining a classical optimization loop with additional steps to optimize the number of kernels---and therefore the memory cost---of our representation of the actual solution to equation~\ref{eqn:rendering}. The remainder of this section describes our function space and optimization procedure.

\myparagraph{Parameterized radiance space}
We represent $L_\theta$ as a weighted sum of $M$ (variable) $13$D-Gaussians over 
position $\mathbf x$, direction $\omega$, normal $\mathbf n$, material attributes (albedo $a$ and roughness $r$): 
\begin{equation}
	L_\theta(\mathbf x, \omega) = \sum_{k=1}^M c_k e^{-(\mathbf q - \mathbf m_k)^\intercal\Sigma_k^{-1}(\mathbf q - \mathbf m_k)}\label{eqn:gp},
\end{equation}
where $\mathbf q=(\mathbf x, \omega, \mathbf n, a, r) \in \mathbb{R}^{13}$,  $\mathbf m_k$ and $\Sigma_k$ are the mean and covariance matrix of the $k$\textsuperscript{th} Gaussian and $c_k$ is its radiance scaling coefficient. While albedo and positions are naturally 3D, normals and directions are plunged in $\mathbb R^3$, and roughness is unidimensional.
This approach, similar to the one of \citet{diolatzis_n_dimensional_2024}, enables sharing Gaussian contributions between nearby points in the scene even when the points do not belong to one contiguous region. For instance, surfaces with spatially-varying materials could share a Gaussian across regions of identical albedo and roughness. See Section~\ref{sec:implementation} for a discussion on replacing $\Sigma$ with a block-diagonal matrix. 
Note that this set of variables does not limit the range of materials that can be used in a scene, but simply facilitates a better adaptation of the Gaussians to variations in the illumination.
We initialize using 10K Gaussians centered at randomly sampled surface points and directions, with low-intensity colors before optimization.

\myparagraph{Gradient estimation}
The loss in Equation~\ref{eqn:loss} is estimated using Monte Carlo integration and
gradients of the loss with respect to $\theta$ are estimated using stochastic gradient descent.
Surface points $\mathbf x_j$, outgoing (resp. incoming) directions $\omega_j$ (resp. $\omega'_j$) are 
generated following adapted probability density functions $p(\mathbf x,\omega)$.
The gradients of the loss are thereby estimated as:
\begin{equation}
	\nabla_\theta \mathcal{L}(\theta) \approx \frac{1}{N}\sum_{j=1}^N \frac{2r_\theta(\mathbf x_j, \omega_j)\nabla_\theta r_\theta(\mathbf x_j, \omega_j)}{p(\mathbf x_j, \omega_j)}.\label{eqn:gradients}
\end{equation}

Ideally, unbiased gradient estimates would require the inner integrals in $r_\theta(\mathbf x_j, \omega_{j})$ and $\nabla_\theta r_\theta(\mathbf x_j, \omega_{j})$ to be evaluated using independent samples.
Following the semi-gradient formulation of Cho and Cho~\shortcite{cho_fast_2024}, we instead drop the gradient of the inner integral entirely, backpropagating only through $r_\theta(\mathbf x_j, \omega_{j})$, trading a small amount of bias for a
significant reduction in gradient variance.

\bigskip
\myparagraph{Densification and pruning} 
We adapt the density of Gaussians where the residual is large, improving the efficiency of the Gaussian mixture $L_\theta$ via splitting/spawning and pruning steps.

We split Gaussians that exhibit large gradients within specific subspaces since this indicates under-resolved variations~\cite{kerbl_gaussian_2023}. 
We split them along the principal axis of the corresponding subspace.
While splitting refines the existing structure, it cannot introduce primitives in regions with no coverage.
Thus we additionally perform residual-driven spawning:
while computing the residual $r(\mathbf x_j,\omega_j)$ in Equation~\ref{eqn:loss}, we select the top-$k$ samples with the highest contribution, and initialize 
new Gaussians centered at the corresponding locations $\mathbf q(\mathbf x_j, \omega_j)$. Their initial radiance coefficient $c_k$ is set proportional to the estimated scattered radiance at these samples given by the right-hand side of the residual $c_k=E(\mathbf x,\omega)+\mathbf TL_\theta(\mathbf x,\omega)$, 
which naturally estimates $L_\theta(\mathbf x,\omega)$ for subsequent optimization.

We prune Gaussians based on their relative normalized contributions, $r_k$ for each Gaussian $G_k$, w.r.t. the global radiance field $L_\theta$. We restrict the integration domain to the effective support $\mathcal{D}_k$ of each Gaussian, i.e. the region of $\mathcal{S} \times \Omega$ where the exponent of $G_k$ exceeds a minimum value $\beta=-14$:
\begin{equation}
    r_k = \int_{\mathcal{D}_k} \frac{c_k G_k(\mathbf q(\mathbf x, \omega) - \mathbf m_k)}{L_\theta(\mathbf x, \omega)} \, d\mathbf x \, d\omega.
\end{equation}
We estimate this integral via the Monte-Carlo estimator
\begin{equation}
    r_k \approx \frac{1}{N} \sum_{j=1}^N \frac{c_k G_k(\mathbf q(\mathbf x_j, \omega_j) - \mathbf m_k)}{L_\theta(\mathbf x_j, \omega_j)}.
\end{equation}
Gaussian kernels whose contribution $r_k$ falls below a threshold $\tau$ are considered negligible and are removed from the representation.
We use $\tau=8 \times 10^{-3}$ for all experiments.
Thus, we prune based on relative energy contribution rather than absolute magnitude, allowing both bright and low-intensity regions to be treated consistently.

\subsection{Model evaluation and rendering} \label{evaluation}

We produce images from the learned model using two alternative ways: (1) \texttt{GS} which simply evaluates $L_\theta$ for every visible point $\mathbf x$ in the scene along directions $\omega$ pointing toward the camera; or (2) \texttt{GS\,+\,MC}, where one bounce of rays are traced from the camera and the right-hand side of Equation~\ref{eqn:kajiya} is evaluated by integrating the incident illumination $L_i$ at this point using
$L_i(\mathbf x,\omega')=L_\theta(\mathbf y,\omega'')$ denoting $(\mathbf y,\omega'')$ as the coordinates of the point seen from $\mathbf x$ in direction $\omega'$.

Both optimization and rendering queries necessitate the evaluation of contributions from potentially large numbers of Gaussian kernels. While queries generated during \texttt{GS} rendering exhibit strong spatial and directional coherence, this is not the case during training or when rendering with \texttt{GS\,+\,MC}, 
where samples are drawn randomly over both position and direction, which precludes screen-space tiling strategies, such as those used in~\citet{kerbl_gaussian_2023}.

For efficient evaluation of $L_\theta$, we introduce a simple yet effective world space tile-based culling strategy that drastically 
reduces the number of Gaussian functions to evaluate per query. Each query is assigned a 64-bit Morton code based on the index of the shape it lies on 
and its 3D position. Queries are then sorted according to their Morton codes and grouped into tiles by truncating the code to a fixed prefix length.
This grouping ensures spatial locality while improving memory coherence during evaluation.
For each tile, we compute conservative bounding ranges across all $13$ input dimensions. These bounds are obtained via parallel reductions over the 
queries assigned to each tile and are used to perform Gaussian/tile intersection tests to cull irrelevant primitives.
If the expected exponent falls below a predefined threshold, the Gaussian contribution to the tile is considered negligible and is discarded, effectively truncating each kernel to a finite region of support.
This hierarchical culling strategy drastically reduces the number of Gaussian kernel evaluations per query. Typically on the \textsc{Bedroom} scene this strategy allows us to 
only evaluate 76 Gaussian kernels on average per pixel, among a total of 22K.

\begin{figure}[tb]
	\centering
	\includegraphics[width=.48\textwidth]{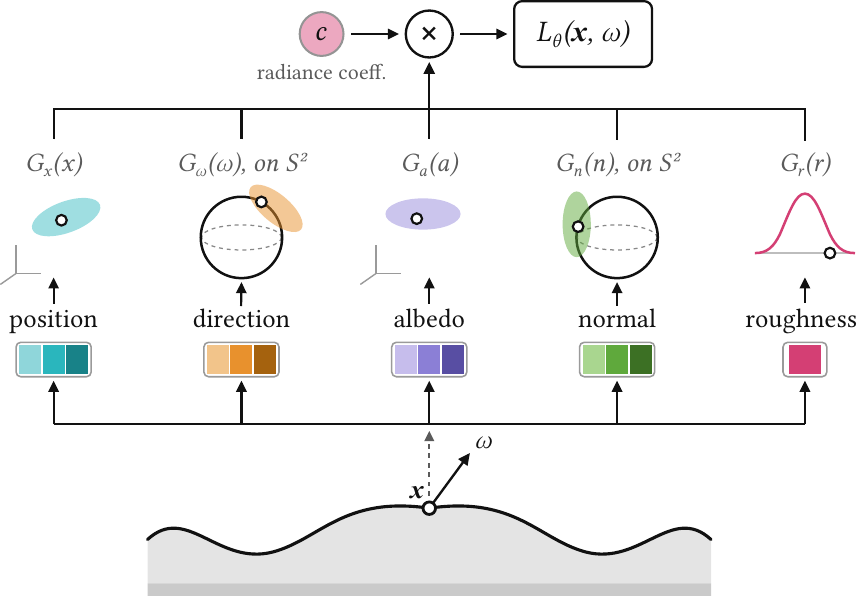}
	\caption{A visualization of our Gaussian representation. Albedo, normal and roughness are extracted from query point $\mathbf{x}$ and serve as additional selection features for the model. Note that Gaussians in albedo and normal have no rotation because of their diagonal representation.}
	\label{fig:gaussian_eval}
\end{figure}

\section{Implementation}\label{sec:implementation}

We implemented our optimization and rendering steps within the Mitsuba / Python framework, which provides easy management of scenes and lighting primitives~\cite{jakob_mitsuba3_2022}.
The Gaussian representation and training pipeline are implemented using PyTorch.
To achieve high efficiency during both training and inference, model evaluation and analytic derivative computations are implemented through custom Taichi kernels~\cite{hu_taichi_2019}.

\bigskip\myparagraph{Covariance matrices}
In the na\"{\i}ve version of our algorithm, covariance matrices are represented using their Cholesky decomposition $\Sigma = \mathbf L^\intercal \mathbf L$,
where $\mathbf L$ is a lower $13\times 13$ triangular matrix. With this choice we have
	$$ \mathbf v^\intercal\mathbf \Sigma^{-1}\mathbf v
		= \|\mathbf L^{-1}\mathbf v \|^2,$$
which means that computing each Gaussian term in Equation~\ref{eqn:gp} boils down to solving a single $13$D triangular linear system.
We found that an interesting trade-off is to turn the full 13-dimensional Gaussian representations into a product of lower-dimensional
Gaussians, thereby neglecting the possible correlation between some of the dimensions. Each Gaussian $G_k$ is modeled as a product of independent Gaussians over these subspaces.
Although this factorization limits the ability to capture cross-subspace correlations, we find that it has a relatively low impact on the quality of the 
converged models, while significantly improving performance by maintaining spatial locality in 3D. 
In all results shown in this paper, position, direction, material and normals are treated as separate subspaces.
Following \citet{kerbl_gaussian_2023}, the covariance matrices of the position and direction components
are decomposed into a scaling matrix $\mathbf S$ and a rotation matrix $\mathbf R$: 
			$$\Sigma = \mathbf R\mathbf S\mathbf S^\intercal \mathbf R^\intercal, $$
making the evaluation of $\mathbf v^\intercal \Sigma^{-1}\mathbf v$ even faster.

Gaussian colors and covariance scales both use an exponential activation, since radiance and scale values lie in the range $[0, \infty)$.

\bigskip\myparagraph{Loss function}
Because radiance is optimized in linear space, applying the na\"{\i}ve loss function of Equation~\ref{eqn:loss} would prioritize learning bright regions 
at the expense of darker ones. To mitigate this, we found very effective to normalize the loss by the current prediction of the model: 
\begin{equation}
	\mathcal{L}(\theta)= \left\Vert \frac{ r_\theta }{\nabla_0(L_\theta) + \epsilon} \right\Vert ^2,
\end{equation}
where $\nabla_0(L_\theta)$ is in effect $L_\theta$, but considered constant when differentiating.

\begin{figure}
	\centering
	\includegraphics[width=\linewidth]{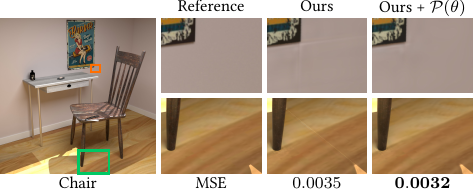}
	\caption{Adding the extra penalty term $\mathcal P(\theta)$ in the loss to control the anisotropy of Gaussian functions allows to avoid
		artifacts due to overly elongated Gaussians such as the thin bright stripe in the middle inset.}
	\label{fig:anisoloss}
\end{figure}

Additionally, we observed that regions containing near-coincident surfaces---such as object contact areas or thin geometric interfaces---present sharp discontinuities that the optimization tries to resolve by strongly elongating the Gaussian kernels along their positional components.
We mitigate this by introducing an anisotropy penalty term $\mathcal P(\theta)$ in the loss function for the three positional components.
Let $\mathbf{s}_i \in \mathbb{R}^3$ denote the logarithmic scaling parameters of Gaussian $i$, with components
$\mathbf{s}_i = (s_{i,1}, s_{i,2}, s_{i,3})$.
We define the mean logarithmic scale as
\begin{equation}
\bar{s}_i = \frac{1}{3}\sum_{n=1}^{3} s_{i,n}.
\end{equation}
The anisotropy regularization term is defined over all Gaussians as
\begin{equation}
	\mathcal{P}(\theta) = \frac{1}{N} \sum_{i=1}^{N} \mathrm{ReLU} \left( \sum_{n=1}^{3} (s_{i,n} - \bar{s}_i)^2 - \epsilon \right),
\end{equation}
where $\epsilon$ controls the tolerated level of anisotropy. In our implementation, we use $\epsilon = 2.5$. Figure~\ref{fig:anisoloss} illustrates the visual 
improvement brought by this anisotropy regularization term.

In summary, the full expression of the loss function we are using during the optimization step is
		$$ \mathcal{L}(\theta)= \left\Vert \frac{ r_\theta }{\nabla_0(L_\theta) + \epsilon} \right\Vert ^2 + \lambda_{\mathrm{aniso}}\mathcal{P}(\theta),$$
where $\lambda_{\mathrm{aniso}}$ monitors the compromise between the shape of the Gaussian in position and the resulting fit of Equation~\ref{eqn:kajiya}. We used $\lambda_{\mathrm{aniso}} = 0.008$ in 
all our examples.

\bigskip\myparagraph{Optimization}
During the estimate of gradients (Equation~\ref{eqn:gradients}) we sample $Q$ incident directions and $Q$ explicit light samples in $r_\theta$ for each outgoing direction $\omega_j$, so as to 
reduce variance due to the generally significant contribution of direct illumination. In all our examples, we used $Q=32$.

Since the evaluation of $r_\theta$ requires $L_\theta$ to be computed twice, we concatenate the queries of both evaluations into a single tensor that is evaluated once, making the economy of
one spatial reordering and culling per iteration. 
Additionally, culling is reused during the backward pass to speed up backpropagation.

During the Gaussian splitting phase, each candidate Gaussian is split along a single subspace at a time with priority given to spatial dimensions, in order to ensure that the child Gaussians accurately 
approximate their parent, and to preserve continuity during optimization. The child Gaussians are initialized by offsetting them in opposite directions along the principal axis of the parent, at a 
distance $d=0.5$ times the principal scale, and scaling their covariance by a factor $k=0.8$ along this axis.

We use the Adam optimizer~\cite{kingma_adam_2015}.
All runs use the same hyperparameters, and learning rates for ($\mathbf{m}$, 
$\mathbf{R}$, $\mathbf{S}$, $c$) are set to
($3\times10^{-4}$, $1\times10^{-3}$, $5\times10^{-3}$, $1\times10^{-2}$).
Gaussians are split every 400 iterations, and 500 new Gaussians are spawned every 500 iterations.
Both splitting and spawning stop at iteration 15000. Gaussians
are pruned every 2000 iterations throughout training.

\bigskip\myparagraph{Rendering}
We exploit the separability of our Gaussian functions to derive an efficient separable culling criterion.
Specifically, we evaluate a Mahalanobis distance between the Gaussian and the tile bounds independently in each subspace.
Because the Gaussian exponent decomposes additively across decorrelated subspaces, these per-subspace distances can be accumulated to form a global bound.
We first apply this test to the 1D subspaces, which provides a cheap and effective early rejection mechanism,
then a similar test is performed for the 3D spatial and directional components using the tile's bounding box.
For 3D subspaces, a good conservative Mahalanobis distance between a tile bounding box and a Gaussian can be computed as follows:
we first compute the Mahalanobis distance from the Gaussian to the center of the bounding box, from which we subtract the maximum distance from the 
center of the bounding box to one of its corners.  Similarly to Hadadan et al.~\shortcite{hadadan_neural_2021}, the model does not directly learn light distributions 
over perfectly specular surfaces. Instead, during both rendering and optimization, paths are extended until the first smooth surface.

\section{Discussion}

Our representation shares similarities with N-dimensional Gaussians \cite{diolatzis_n_dimensional_2024}: by centering Gaussians on scene features (albedo, normals), high-frequency content is encoded directly in the input space rather than learned.
Their method fits pre-generated path-traced images, solving transport beforehand, whereas we solve the rendering equation itself.
We keep fewer principal component axes than their full representation (see Section~\ref{sec:implementation}).
Dropping this generality allows us to derive a more efficient culling strategy well suited for real-time rendering, while reducing the memory footprint of our representation.

Our method belongs to the larger family of kernel-based approximations. For instance, our adaptive Gaussian mixture is equivalent 
to constructing a Gaussian process with a location-dependent covariance function, i.e. a nonstationary 
kernel where the notion of distance changes across space.
Similarly, neural network approximations (e.g. Neural Radiosity~\cite{hadadan_neural_2021}) implicitly define a kernel, but use fixed positions on 
a predefined grid. As a comparison, our model not only learns kernel weights but also positions $\mathbf m_k$ in $13$D, shape (covariance matrices $\Sigma_k$) and number 
of nodes to use.  Finally, our model can also be understood as a learned radial basis function neural network, that implicitly constructs a nonstationary 
Gaussian process kernel. In different domains Gaussian processes and kernel methods have been applied to solve partial differential 
equations~\cite{chen_solving_2021,xu_toward_2025}, while sparsity-promoting techniques improve efficiency~\cite{chaudhuri_improved_2022}.

One advantage of our adaptive representation is that it avoids the pitfalls of bandwidth-limited constructions~\cite{soler_theoretical_2022}.

Since $L_\theta$ appears twice in the expression of the residual $r_\theta$, the model must be evaluated twice: for outgoing radiance, and at the intersection 
points for incident radiance. This can be interpreted numerically as a recursive training approach, where the model iteratively 
uses its own previous state as the best estimate so far to the solution of the rendering equation.

\myparagraph{Limitations} 
	 The calculation of $r_\theta$ is currently conducted using uniform sampling of surfaces, which favors the learning of Gaussians with a large positional footprint. 
		We noticed that glossy reflections on curved surfaces are learned quicker with a sampling that is based on curvature. We think that sampling according to 
		a local prediction of angular and spatial bandwidth~\cite{belcour_covariance_2013} would eventually focus the workload on regions that need it.

		Our current implementation runs on static scenes only. We think that if single objects are moved or scene parameters such as reflectance properties or lighting are changed,
		it would be highly beneficial to start from the previous solution and continue optimizing from there.
		Another possible avenue for predefined changes would be to add an extra temporal dimension to Gaussian functions.

		Our culling approach relies on the assumption that queries grouped within a tile are coherent both spatially and angularly.
		While this assumption holds well for primary camera rays during \texttt{GS} rendering, it becomes weaker during training and \texttt{GS\,+\,MC} rendering, where queries exhibit reduced spatio-directional correlation.
		Consequently, directional culling becomes less effective, resulting in a 1.5$\times$ inference slowdown in the worst case.
		Future work could investigate direction-aware tiling or clustering schemes to better preserve spatio-directional locality under these workloads.

		Like any Gaussian-based representation, ours is smooth by construction and resolves a
		discontinuity only where an input announces it. Albedo, normal and roughness jump
		at material and geometric edges, so boundaries aligned with them stay sharp
		(Fig.~\ref{fig:material}). Position and direction vary continuously, so radiance
		discontinuous in those alone is interpolated and softened.
		Near-specular reflections show it most: the chimney's dark base in
		\textsc{Living Room}, and the bright highlight along the seat edge in \textsc{Chair} (Fig.~\ref{fig:modes}).

\section{Results}

A typical example of exploring the solution to the light transport equation in a complex scene showing glossy and specular materials, complex geometry and textures is shown in 
Figure~\ref{fig:teaser}. Once the solution is optimized, all three images are obtained by simply slicing the Gaussians w.r.t. the scene geometry and materials, and the camera direction. For each viewpoint we show the rendered vs. flattened Gaussians, the directional component of which makes their visualized support ellipsoids move along with the camera.

\begin{figure}[t]
  \centering
  \includegraphics[width=\columnwidth]{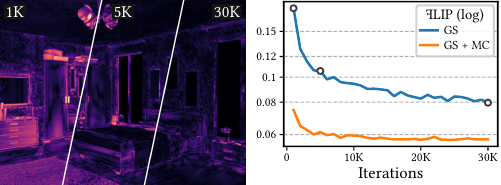}
  \caption{%
    \emph{Left:} \reflectbox{F}LIP error maps of \textsc{Bedroom}, each computed from a different checkpoint of the same run.
	Brighter is worse.
    \emph{Right:} mean \reflectbox{F}LIP against training iteration for both render modes at \num{1920}$\times$\num{1080} \num{64}~spp.
  }
  \label{fig:convergence}
\end{figure}

Figure~\ref{fig:convergence} illustrates the convergence of our method.
We use the HDR-\reflectbox{F}LIP error metric \cite{andersson_visualizing_2021} for quantitative comparisons.
Although the limited total number of Gaussians used to represent the solution inherently
biases the result, visual and \reflectbox{F}LIP comparisons with the path traced reference prove very satisfactory.

\begin{table*}[t]
    \centering
    \small
     \caption{
       MSE comparison of ablation runs (lower is better).
       When the full model is not ranked first (e.g., \textsc{Rings}), the performance
       gap is marginal and within run-to-run stochastic variance.
       Colors mark the {\setlength{\fboxsep}{2pt}\colorbox{best}{\textbf{best}}},
       {\setlength{\fboxsep}{2pt}\colorbox{second}{second-best}}, and {\setlength{\fboxsep}{2pt}\colorbox{third}{third-best}}
       result per column.
     }
    \sisetup{table-format=1.5, table-number-alignment=center, detect-weight, group-digits=false}
    \begin{tabular}{|l | *{7}{S}|}
        \hline
        \textbf{Ablation} & {\textsc{Bedroom}} & {\textsc{Chair}} & {\textsc{Dining Room}} & {\textsc{Living Room}} & {\textsc{Rings}} & {\textsc{Staircase}} & {\textsc{Veach Ajar}} \\
        \hline

        No material (9D) & 
        0.02313 & 
        0.00188 & 
        {\cellcolor{third}0.00205} & 
        {\cellcolor{second}0.02368} & 
        0.02001 & 
        0.00759 & 
        0.01440 \\

        No normal (10D) & 
        {\cellcolor{second}0.01137} & 
        0.00215 & 
        0.00259 & 
        0.04286 & 
        0.02005 & 
        0.00463 & 
        0.01179 \\

        No Split & 
        0.03354 & 
        0.00348 & 
        0.00324 & 
        0.06272 & 
        0.02016 & 
        0.00353 & 
        0.02557 \\

        No Spawn & 
        0.01503 & 
        {\cellcolor{third}0.00167} & 
        {\cellcolor{second}0.00204} & 
        0.03452 & 
        {\cellcolor{second}0.02001} & 
        {\cellcolor{third}0.00239} & 
        {\cellcolor{third}0.00591} \\

        No $\mathcal{P}(\theta)$ & 
        {\cellcolor{third}0.01339} & 
        {\cellcolor{second}0.00130} & 
        0.00207 & 
        {\cellcolor{third}0.02820} & 
        {\cellcolor{best}\textbf{0.02000}} & 
        {\cellcolor{best}\textbf{0.00126}} & 
        {\cellcolor{second}0.00538} \\

        Full & 
        {\cellcolor{best}\textbf{0.00636}} & 
        {\cellcolor{best}\textbf{0.00116}} & 
        {\cellcolor{best}\textbf{0.00189}} & 
        {\cellcolor{best}\textbf{0.02203}} & 
        {\cellcolor{third}0.02001} & 
        {\cellcolor{second}0.00131} & 
        {\cellcolor{best}\textbf{0.00444}} \\

        \hline
    \end{tabular}
    \label{tab:ablation}
\end{table*}

\begin{figure}[tbp]
	\centering
	\includegraphics[width=\linewidth]{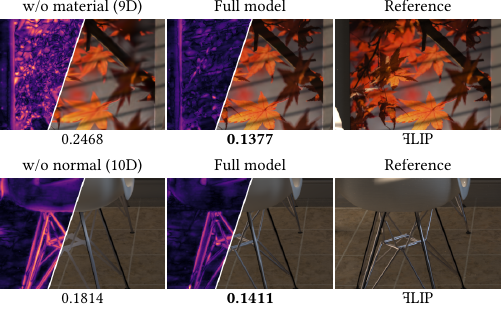}
	\caption{The extra dimensions of the Gaussians allow accommodation of local variation. For example, the Gaussian visualizations show that not including albedo in the Gaussian dimensions requires more (scene total 37.9K) Gaussians in textured regions than with albedo (scene total 35.2K Gaussians).}
	\label{fig:material}
\end{figure}

In Table~\ref{tab:ablation} we present an ablation study that illustrates the relative impact of Gaussian dimensionality, spawning, splitting and regularization term on the mean square error for all scenes presented in this paper. Except for the \textsc{Rings} scene where these stages have little to no impact, the improvement brought by
these steps in terms of image quality stands out clearly. When splitting is disabled, as it is normally the main source of added detail, we increase the number 
of initial Gaussians to 30K for fair comparison. In Figure~\ref{fig:material} we illustrate the visual impact of not including material and normal dimensions within the 
$13$ dimensions of the space of Gaussians.

We compare against Neural Radiosity \cite{hadadan_neural_2021}, extended with two feature encodings: the multi-resolution hash grid of \citet{muller_instant_2022} and the vertex features of \citet{su_vertex_2025}.
We use the official implementation of Su et al. with their default settings (four learnable parameters per vertex) and adapt it to support hash grids:
eight resolution levels, a coarsest resolution of $4$, a scale factor of $2$ between successive levels, and eight learnable parameters per voxel.
Both baselines are trained for the same number of iterations and with the same network
(four hidden layers of $256$ neurons) so that they differ only in how the input is encoded.
Training and rendering are performed on an NVIDIA RTX~4080~SUPER for all methods.

Table~\ref{tab:training_stats} reports training and rendering cost against the stronger of the two baselines, vertex features.
Our models train for a fixed \num{30}K iterations (\num{7}--\num{12} minutes), whereas the baselines were trained until their
artifacts cleared, \num{40}K--\num{100}K iterations (\num{1.5}--\num{4.5} hours): a \num{10.0}--\num{23.4}$\times$ reduction.
At one sample per pixel we render \num{2.2}--\num{4.5}$\times$ faster, and reach a lower \reflectbox{F}LIP on five of the six scenes.
\begin{table}[t]
    \centering
    \small
    \caption{
      Training and rendering cost against Neural Radiosity with vertex
      features, NR~(VF). Model is the size of the trained representation on disk.
      Render is the frame time at $1920\times1080$, 1~spp.
    }
    \setlength{\tabcolsep}{1pt}%
    \begin{tabular*}{\columnwidth}{@{\extracolsep{\fill}}c<{\hspace{2.878pt}}>{\hspace{2.878pt}}c<{\hspace{2.878pt}}>{\hspace{2.878pt}}c<{\hspace{2.878pt}}>{\hspace{2.878pt}}r<{\hspace{2.878pt}}>{\hspace{2.878pt}}r<{\hspace{2.878pt}}>{\hspace{2.878pt}}r<{\hspace{2.878pt}}>{\hspace{2.878pt}}r@{}}
        \toprule
        \textbf{Scene} & \textbf{Method} & \textbf{Iter.} & \textbf{Train} & \textbf{Model} & \textbf{Render} & \textbf{\reflectbox{F}LIP} \\
        \midrule
        \rowcolor[gray]{0.92}[2.00pt][2.00pt] & NR (VF) & 40K & 2h10m & 26.4\,MB & 70.4\,ms & \textbf{0.0774} \\
        \rowcolor[gray]{0.92}[2.00pt][2.00pt]\multirow{-2}{*}{\textsc{Bedroom}} & Ours & 30K & 10m48s & 4.4\,MB & 18.1\,ms & 0.0824 \\
        \addlinespace[2pt]
         & NR (VF) & 40K & 1h42m & 1.9\,MB & 52.3\,ms & 0.0602 \\
        \multirow{-2}{*}{\textsc{Chair}} & Ours & 30K & 10m07s & 4.3\,MB & 16.1\,ms & \textbf{0.0486} \\
        \addlinespace[2pt]
        \rowcolor[gray]{0.92}[2.00pt][2.00pt] & NR (VF) & 40K & 1h54m & 6.0\,MB & 60.2\,ms & 0.1012 \\
        \rowcolor[gray]{0.92}[2.00pt][2.00pt]\multirow{-2}{*}{\textsc{Dining Room}} & Ours & 30K & 9m17s & 5.4\,MB & 13.5\,ms & \textbf{0.0831} \\
        \addlinespace[2pt]
         & NR (VF) & 100K & 4h29m & 5.3\,MB & 61.5\,ms & 0.1362 \\
        \multirow{-2}{*}{\textsc{Living Room}} & Ours & 30K & 11m30s & 5.8\,MB & 16.5\,ms & \textbf{0.1177} \\
        \addlinespace[2pt]
        \rowcolor[gray]{0.92}[2.00pt][2.00pt] & NR (VF) & 40K & 1h30m & 0.9\,MB & 44.0\,ms & 0.0509 \\
        \rowcolor[gray]{0.92}[2.00pt][2.00pt]\multirow{-2}{*}{\textsc{Rings}} & Ours & 30K & 7m37s & 1.0\,MB & 19.9\,ms & \textbf{0.0425} \\
        \addlinespace[2pt]
         & NR (VF) & 60K & 2h25m & 6.5\,MB & 46.5\,ms & 0.1499 \\
        \multirow{-2}{*}{\textsc{Veach Ajar}} & Ours & 30K & 12m08s & 4.1\,MB & 13.0\,ms & \textbf{0.0641} \\
        \midrule
         & NR (VF) & 53K & 2h22m & 7.8\,MB & 55.8\,ms & 0.0960 \\
        \multirow{-2}{*}{Average} & Ours & 30K & 10m15s & 4.2\,MB & 16.2\,ms & \textbf{0.0730} \\
        \bottomrule
    \end{tabular*}
    \label{tab:training_stats}
\end{table}

Figure~\ref{fig:comparisonNR} provides a visual comparison of the three methods at equal sample count.
Additional viewpoints in \textsc{Living Room} are provided at 1~spp in Figure~\ref{fig:teaser}.
Model size (the trained representation as stored on disk) behaves differently for the two encodings:
the hash grid is a fixed \num{35.7}\,MB irrespective of the scene, while vertex features mostly scale with the vertex count of the geometry and range from \num{0.9}\,MB to \num{26.4}\,MB.
Our representation is bounded by the number of Gaussians rather than by either grid resolution or tessellation, and stays between \num{1}\,MB and \num{5.9}\,MB on all scenes.
The peak GPU memory of the process---including the renderer's own allocations---is \num{3.1}--\num{4.3}\,GB during training and \num{3.0}--\num{4.0}\,GB when
rendering in \texttt{GS}, rising to \num{3.9}--\num{6.1}\,GB in \texttt{GS\,+\,MC}, regardless of the sample count.

Table~\ref{tab:modes} connects rendering mode, sample count, error, and frame time for our method.
Since \texttt{GS} mode queries the representation at the primary hit, samples only buy antialiasing and noise reduction on traced specular paths.
\texttt{GS\,+\,MC} takes one Monte Carlo bounce first, and so is noisier at one sample and more accurate once converged, overtaking \texttt{GS} at \num{1024} samples on every scene. 
\begin{table}[t]
    \centering
    \small
    \caption{
      Rendering mode vs. sample count for our method.
      Eval is the model evaluation alone, so the gap to 1~spp frame time is renderer overhead.
      On \textsc{Rings} secondary rays leave the scene almost everywhere, so \texttt{GS\,+\,MC} evaluates the model far less than \texttt{GS} and ends up cheaper.
    }
    \setlength{\tabcolsep}{1pt}%
    \begin{tabular*}{\columnwidth}{@{\extracolsep{\fill}}l<{\hspace{0.554pt}}>{\hspace{0.554pt}}l<{\hspace{0.554pt}}>{\hspace{1.554pt}}r<{\hspace{0.554pt}}>{\hspace{0.554pt}}r<{\hspace{0.554pt}}>{\hspace{0.554pt}}r<{\hspace{1.554pt}}>{\hspace{1.554pt}}r<{\hspace{0.554pt}}>{\hspace{0.554pt}}r<{\hspace{1.554pt}}>{\hspace{1.554pt}}r<{\hspace{0.554pt}}>{\hspace{0.554pt}}r@{}}
        \toprule
         &  & \multicolumn{3}{c}{\textbf{1 spp}} & \multicolumn{2}{c}{\textbf{64 spp}} & \multicolumn{2}{c}{\textbf{1024 spp}} \\
        \cmidrule(lr){3-5} \cmidrule(lr){6-7} \cmidrule(lr){8-9}
        \textbf{Scene} & \textbf{Mode} & \textbf{Eval} & \textbf{Render} & \textbf{\reflectbox{F}LIP} & \textbf{Render} & \textbf{\reflectbox{F}LIP} & \textbf{Render} & \textbf{\reflectbox{F}LIP} \\
        \midrule
        \rowcolor[gray]{0.92}[1.88pt][1.88pt]\textsc{Bed-} & \texttt{GS} & 10.5\,ms & 18.1\,ms & \textbf{0.101} & 1.15\,s & 0.082 & 18.03\,s & 0.082 \\
        \rowcolor[gray]{0.92}[1.88pt][1.88pt]\textsc{room} & \texttt{GS\,+\,MC} & 14.3\,ms & 31.1\,ms & 0.276 & 1.88\,s & \textbf{0.058} & 30.02\,s & \textbf{0.027} \\
        \addlinespace[2pt]
         & \texttt{GS} & 11.2\,ms & 16.1\,ms & \textbf{0.052} & 0.99\,s & \textbf{0.049} & 15.44\,s & 0.048 \\
        \multirow{-2}{*}{\textsc{Chair}} & \texttt{GS\,+\,MC} & 11.8\,ms & 19.3\,ms & 0.329 & 1.23\,s & 0.064 & 19.71\,s & \textbf{0.025} \\
        \addlinespace[2pt]
        \rowcolor[gray]{0.92}[1.88pt][1.88pt]\textsc{Dining} & \texttt{GS} & 10.2\,ms & 13.5\,ms & \textbf{0.090} & 0.86\,s & \textbf{0.083} & 13.09\,s & 0.083 \\
        \rowcolor[gray]{0.92}[1.88pt][1.88pt]\textsc{Room} & \texttt{GS\,+\,MC} & 14.6\,ms & 21.2\,ms & 0.442 & 1.36\,s & 0.106 & 21.66\,s & \textbf{0.041} \\
        \addlinespace[2pt]
        \textsc{Living} & \texttt{GS} & 10.6\,ms & 16.5\,ms & \textbf{0.125} & 1.02\,s & \textbf{0.118} & 16.26\,s & 0.117 \\
        \textsc{Room} & \texttt{GS\,+\,MC} & 13.9\,ms & 26.3\,ms & 0.547 & 1.67\,s & 0.140 & 26.51\,s & \textbf{0.056} \\
        \addlinespace[2pt]
        \rowcolor[gray]{0.92}[1.88pt][1.88pt] & \texttt{GS} & 14.2\,ms & 19.9\,ms & \textbf{0.062} & 1.28\,s & \textbf{0.042} & 20.04\,s & 0.039 \\
        \rowcolor[gray]{0.92}[1.88pt][1.88pt]\multirow{-2}{*}{\textsc{Rings}} & \texttt{GS\,+\,MC} & 3.8\,ms & 14.7\,ms & 0.389 & 0.94\,s & 0.074 & 14.81\,s & \textbf{0.027} \\
        \addlinespace[2pt]
        \textsc{Veach} & \texttt{GS} & 10.1\,ms & 13.0\,ms & \textbf{0.070} & 0.82\,s & \textbf{0.064} & 12.83\,s & 0.064 \\
        \textsc{Ajar} & \texttt{GS\,+\,MC} & 14.2\,ms & 19.4\,ms & 0.350 & 1.24\,s & 0.094 & 19.70\,s & \textbf{0.038} \\
        \bottomrule
    \end{tabular*}
    \label{tab:modes}
\end{table}

Finally, in Figure~\ref{fig:modes} we show that our method can successfully handle very different situations typical of global illumination: geometric complexity, 
caustics, glossy and specular surfaces, textures. Gaussian components naturally decompose into diffuse + glossy components when BSDFs contain multiple lobes.

\section{Conclusion}

We presented a novel method for solving the rendering equation based on a mixture of Gaussian functions, which allows real-time exploration of
the solution to the problem of global illumination in static scenes. The very low memory footprint of our representation makes it
easy to embed in a viewer for instance to explore architectural constructions.

Our plans for future work include extending our method to the integro-differential equation for global illumination in participating media,
and the export of light fields.
The intrinsic local support of our representation in frequency space also naturally allows to isolate/scale/cancel parts of the solution to
the light transport equation by their corresponding spatial and angular frequency. This could for instance be used to box-filter the signal 
as a post-process for artistic reasons. More generally our model could be used for baking global illumination thanks to its very economic memory footprint.

\section*{Acknowledgments}
We thank the anonymous reviewers for their careful reading of our manuscript and for their
valuable feedback.

We are also grateful to the artists whose scenes we use.
\textsc{Bedroom} by SlykDrako and \textsc{Veach Ajar} by Benedikt Bitterli are released under CC0.
\textsc{Dining Room}, \textsc{Living Room} and \textsc{Staircase} are by Wig42 and are used under CC BY (\url{https://creativecommons.org/licenses/by/3.0/}).
The Mitsuba versions of these five scenes are by Benedikt Bitterli.
\textsc{Chair} and \textsc{Rings} are courtesy of \citet{hadadan_neural_2021}, who adapted
them from scenes by BlendSwap users barcin and IAMCUBEMAN, both used under CC BY.

\appendix

\bibliographystyle{ACM-Reference-Format}
\bibliography{project}

\begin{figure*}
  \centering
  \includegraphics[width=\textwidth]{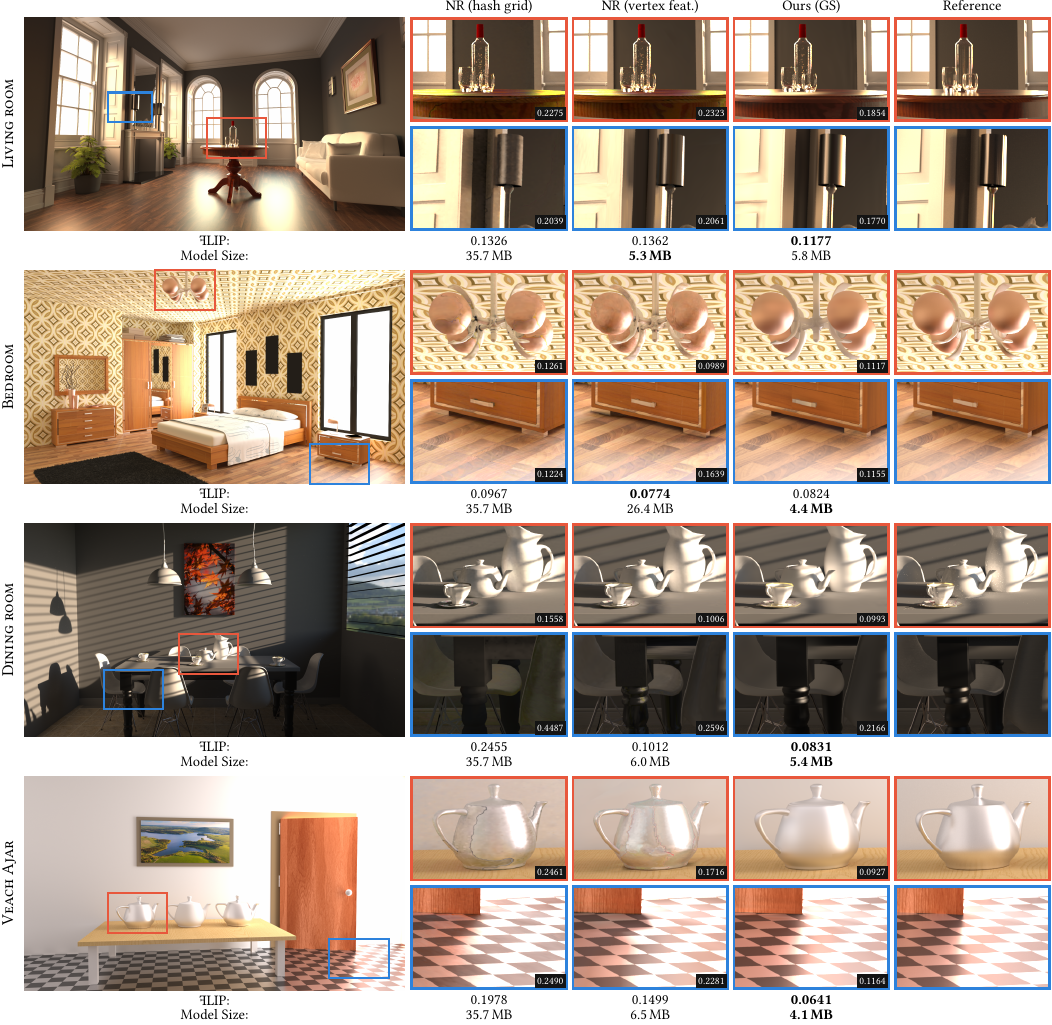}
  \caption{Visual comparison with Neural Radiosity~\cite{hadadan_neural_2021} extended with Hash Grids \cite{muller_instant_2022} and Neural Vertex Features \cite{su_vertex_2025} on multiple test scenes.
  All renders have a resolution of 1920$\times$1080 and use 64~spp. See Table~\ref{tab:training_stats} for training and rendering times.}
	\label{fig:comparisonNR}
\end{figure*}

\begin{figure*}[p]
  \centering
  \includegraphics[width=\textwidth]{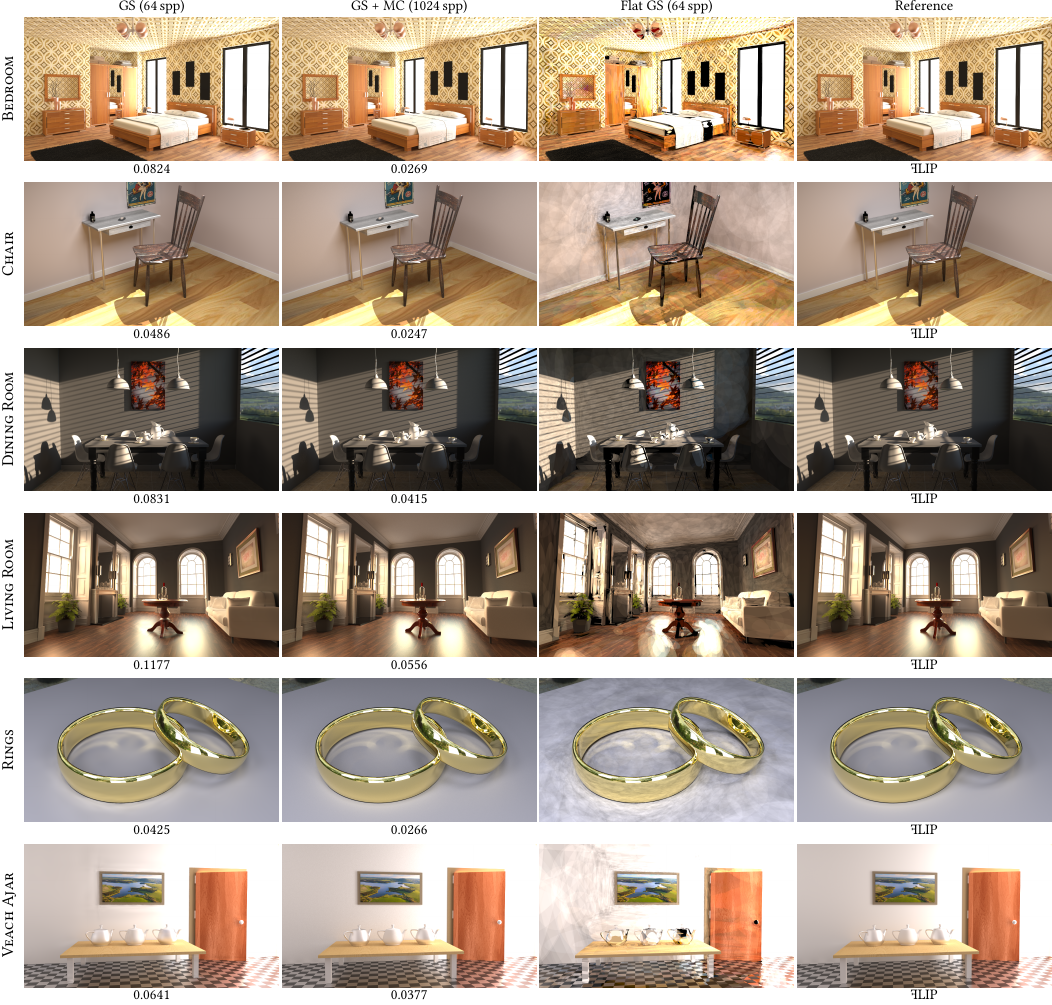}
  \caption{ At the cost of some precomputation, our method allows fast multi-view rendering, with low \reflectbox{F}LIP, of challenging light transport settings such as indirect illumination, caustics, high-frequency shadows and specular reflections.
  In all situations, our \texttt{GS} render (column a.) rapidly approximates the reference.
  In scenes with strong indirect lighting, our \texttt{GS\,+\,MC} render (column b.) improves the quality of the result at 1024~spp.
  The flat-shaded Gaussians (column c.) show adaptation of covariances to the geometry, shadows, etc.
  See Table~\ref{tab:training_stats} for computation times.  
}
  \label{fig:modes}
\end{figure*}

\end{document}